\pdfoutput=1
\documentclass[aps,floats,floatfix,showpacs,amssymb,prd,twocolumn,superscriptaddress,nofootinbib,nolongbibliography,reprint,preprintnumbers]{revtex4-1}

\usepackage{amssymb,amsmath,verbatim,mathtools,needspace,enumitem,etoolbox,graphicx,physics,microtype,afterpage,xspace,tabularx,lmodern,multirow,bm}
\usepackage{textcomp, gensymb}

\usepackage{braket}

\usepackage{relsize}

\usepackage{dcolumn}

\usepackage[normalem]{ulem}
\usepackage{placeins}
\usepackage{comment}
\usepackage[dvipsnames, usenames, table]{xcolor}
\definecolor{linkcolor}{rgb}{0.0,0.3,0.5}
\usepackage[unicode, colorlinks=true, linkcolor=linkcolor, citecolor=linkcolor, filecolor=linkcolor, urlcolor=linkcolor, linktocpage, breaklinks]{hyperref}
\usepackage[all]{hypcap}
\usepackage[T1]{fontenc}
\usepackage[utf8]{inputenc}
\usepackage{mathrsfs}
\hypersetup{colorlinks=true,citecolor=romared,linkcolor=romared,urlcolor=romared}

\definecolor{romared}{RGB}{142,0,28}

\newcommand{\be}{\begin{equation}}
\newcommand{\ee}{\end{equation}}

\def\be{\begin{equation}}
\def\ee{\end{equation}}
\newcommand{\beq}{\begin{eqnarray}}
\newcommand{\eeq}{\end{eqnarray}}

\usepackage{aas_macros}
\usepackage{makecell}
\usepackage{soul}
\usepackage[nolist,nohyperlinks]{acronym}
\usepackage{orcidlink}
\usepackage{lipsum}
\usepackage{booktabs}
\usepackage{bbold}
\usepackage{amssymb,amsmath}
\usepackage{fontawesome}

\acrodef{LSC}[LSC]{LIGO Scientific Collaboration}
\acrodef{BH}{black hole}
\acrodef{NS}{neutron star}
\acrodef{PN}{Post-Newtonian}
\acrodef{BBH}{binary black-hole}
\acrodef{BNS}{binary neutron-star}
\acrodef{NSBH}{neutron-star black-hole}
\acrodef{NR}{numerical relativity}
\acrodef{GW}{gravitational wave}
\acrodef{PSD}{power spectral density}
\acrodef{aLIGO}{Advanced Laser interferometer Gravitational-Wave Observatory}
\acrodef{AZDHP}{aLIGO zero detuned high power density}
\acrodef{GR}{general relativity}
\acrodef{PE}{parameter estimation}
\acrodef{LAL}{LVK algorithm library}
\acrodef{TPI}{tensor-product interpolant}
\acrodef{SVD}{singular value decomposition}
\acrodef{SNR}{signal-to-noise ratio}
\acrodef{ODE}{ordinary differential equation}
\acrodef{PDE}{partial differential equation}
\acrodef{ROM}{reduced order model}
\acrodef{QNM}{quasi-normal mode}
\acrodef{IMR}{inspiral-merger-ringdown}
\acrodef{LVK}{LIGO-Virgo-KAGRA}
\acrodef{SXS}{Simulating eXtreme Spacetimes}

\newcommand{\jhu}{\affiliation{William H. Miller III Department of Physics and Astronomy, Johns Hopkins University, 3400 North Charles Street, Baltimore, Maryland, 21218, USA}}
\newcommand{\ias}{\affiliation{School of Natural Sciences, Institute for Advanced Study, Einstein Drive, Princeton, New Jersey, 08540, USA}}

\newcommand{\NIT}{\affiliation{Department of Electronics and Communication Engineering, National Institute of Technology, Tiruchirappalli 620015, India}}
\newcommand{\UTAustin}{\affiliation{Weinberg Institute, University of Texas at Austin, Austin, TX 78712, USA}}

\graphicspath{{../Plots/}}

\newcommand{\orcid}[1]{\href{https://orcid.org/#1}{\includegraphics[width=10pt]{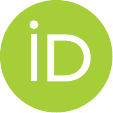}}}
\newcommand*{\rom}[1]{\expandafter\@slowromancap\romannumeral #1@}

\newcommand{\ben}{\begin{enumerate}}
\newcommand{\een}{\end{enumerate}}

\def\be{\begin{equation}}
\def\ee{\end{equation}}
\def\beq{\begin{eqnarray}}
\def\eeq{\end{eqnarray}}

\def\apjl{\rm{ApJ}}
\def\apjs{\rm{ApJS}}
\def\aap{\rm{A\&A}}

\graphicspath{{./Plots/}}

\allowdisplaybreaks

\begin{document}

\pagenumbering{arabic}

\title{Sample-efficient non-Gaussian noise reduction in gravitational wave data via learnable wavelets}

\author{Arush Pimpalkar}
\jhu
\NIT

\author{Digvijay Wadekar}
\jhu
\UTAustin

\author{Mark Ho-Yeuk Cheung}
\jhu
\ias

\author{Emanuele Berti}
\jhu

\date{\today}

\begin{abstract}
We introduce \texttt{WaveletNet}, a wavelet-based neural network architecture to identify and reduce non-Gaussian noise in gravitational wave data.
Traditionally, convolutional neural networks (CNNs) have been widely used as a flexible machine learning method to mitigate non-Gaussian noise. However, training CNNs requires many data samples, especially when the input data segments are long. Glitches that mimic high-mass black hole signals are empirically known to have a wavelet-like structure.
We exploit this property in \texttt{WaveletNet} by using simple neural networks to learn the best family of wavelets to model glitches in the LIGO-Virgo-KAGRA O3 data. Due to its simplicity, our framework is significantly more sample-efficient than CNNs.
As a use case, we build upon the \texttt{TIER} method from Ref.~\cite{Wadekar:2025lhk} and show how \texttt{WaveletNet} can improve the performance of any search pipeline. We take potential GW candidates from the pipeline, and then downweight the candidates having noisy strain regions in their vicinity (i.e., within $t_\mathrm{candidate}\pm \mathcal{O}(10)$ sec). We use our framework in a modular way: we provide an output score which can be added to the pipeline's existing detection statistic score for the candidates.
We test our method using candidates from the \texttt{IAS-HM} search pipeline and show that it improves the search sensitive volume by up to $\sim 15$\% for high-mass, asymmetric binaries. 
\href{https://github.com/Arush-Pimpalkar/WaveletNet}{\faGithub}
\end{abstract}

\maketitle


\section{introduction}\label{sec:intro}
Gravitational waves (GWs), first theorized by Albert Einstein in 1916 as a consequence of his general theory of relativity~\cite{Einstein:1916vd}, are produced by some of the most violent processes in the Universe, such as the inspiral and merger of binary black holes~\cite{2016PhRvL.116f1102A}.

The LIGO, Virgo and KAGRA detectors~\cite{LIGOScientific:2014pky, Cabero:2019orq} are now routinely measuring such merger events. The sensitivity of the detectors is affected by various sources of noise. Some noise sources, called glitches, are non-Gaussian in nature and cannot be eliminated easily. Glitches can imitate real GW strains, making it harder to confirm the astrophysical nature of some of the candidate events~\cite{Cabero:2019orq, Zevin:2016qwy, LIGO:2024kkz, Cabero:2019orq, Son21_Scattered_Light, Cheung:2025grp}. 

There are multiple GW search pipelines that operate in parallel to analyze and detect candidate signals from the interferometer data, each employing different strategies for signal extraction and background estimation. For example, the \texttt{PyCBC} and \texttt{GstLAL} pipelines perform matched filtering with a bank of compact-binary coalescence templates to identify coincident triggers across detectors~\cite{Usman:2015kfa}. They also use a likelihood-ratio based ranking statistic~\cite{Messick:2016aqy}. The coherent WaveBurst (cWB) pipeline, in contrast, is an unmodeled search that reconstructs transient signals through coherent excess-power analysis~\cite{Klimenko:2015ypf}. The Multi-Band Template Analysis (MBTA) pipeline reduces computational cost by splitting the matched filtering into multiple frequency bands~\cite{Aubin:2020goo}. The \texttt{IAS-HM} pipeline enhances standard matched filtering searches by incorporating higher-order harmonic modes beyond the dominant quadrupole, improving sensitivity to asymmetric and high-mass binaries. It performs mode-by-mode filtering and combines the resulting SNRs using physically informed ranking statistics, yielding substantial gains in detection reach.

Low-mass compact binary coalescences produce long-duration signals that spend many waveform cycles in the detectors' sensitive band, where the noise is approximately Gaussian. In contrast, high-mass systems generate short in-band signals that overlap more strongly with non-Gaussian transient disturbances. As a result, the background for high-mass searches is dominated by glitches rather than stationary noise. This non-Gaussian contamination reduces the discriminating power of standard detection statistics and increases the likelihood of misclassifying noise transients as astrophysical events. Consequently, improved data analysis methods are required to distinguish genuine high-mass signals from the transient noise background.

Flexible machine learning (ML) architectures such as CNNs and transformer-based models have been widely applied to the identification of non-Gaussian noise in GW data, see e.g.,~\cite{Koloniari:2024kww, Zhao:2023tqr, Yamamoto:2022adl, Zevin:2023rmt,Gebhard:2019ldz, Li17_CNN, Biswas:2013wfa,Zevin:2016qwy,Cuo24_ML_GW_review, Fernandes:2025yxu, Sharma:2022ibm, Sal24_Deepclean, Rei25_Deepclean_coherence, Alvarez-Lopez_GSpyNet, Lop25_ML_noise, Mal25_PE_glitch, Leg24_PE_glitch, Shah:2023twc, Mar25_Aframe_ML, Schafer:2022dxv,
McL24_ML_search_BNS, Kap17_Classifier, Bin23_ML_CWB, Ess20_iDQ, Nagarajan:2025hws, Cha24_Whisper, Cha24_Aware2, Cha24_Aware3, McI22_signal_consistency_ML}. While these approaches are highly expressive - i.e., they can learn and model complex, nonlinear relationships and hierarchical features in the data - they typically require large training datasets, particularly when operating on long time-domain segments (tens of seconds or more). This data demand can be a limiting factor in regimes where representative examples of specific glitch morphologies are scarce.

Empirically, a class of glitches that contaminate searches for high-mass binary black hole mergers exhibits localized, wavelet-like time-frequency structure. Motivated by this observation, we introduce an explicit inductive bias in our neural network method by modeling glitches using a parameterized wavelet family: the Morlet wavelets ($\psi(t)=e^{-t^2/2\sigma^2}e^{i
\omega_0 t}$), as an approximate generative model. Rather than learning arbitrary features from raw data, we employ a lightweight multi-layer perceptron (MLP) to infer physically interpretable wavelet parameters, such as central frequency and temporal duration. This approach trades architectural flexibility for data efficiency and interpretability, while remaining sufficiently expressive to capture the dominant glitch morphologies relevant to high-mass searches.

\begin{figure*}
    \centering
    \includegraphics[width=\linewidth]{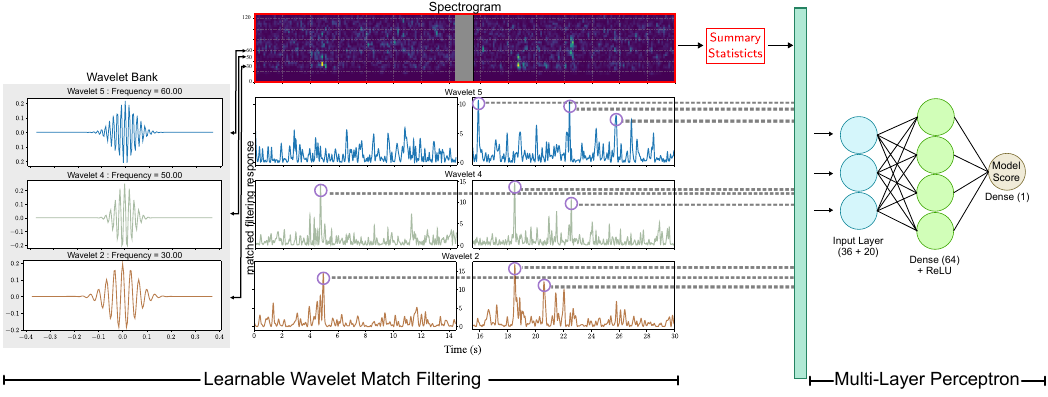}
    \caption{This figure illustrates the \texttt{WaveletNet} model architecture for evaluating the extended strain information in the vicinity of a GW candidate. Extended strain data in the $\pm \sim$15 s region surrounding each candidate is filtered through a bank of learnable Morlet wavelets, acting as matched filters and producing time-domain responses sensitive to transient structure in the environmental data. 
    From each wavelet response, we extract the three largest peaks and their corresponding time indices, yielding a total of 36 wavelet-derived features across the six wavelets that we use. These features are concatenated with 20 additional summary statistical features (see Section~\ref{subsec:nonlocal} for details), forming a combined 36+20=56 dimensional input feature vector. This vector is passed to a multi-layer perceptron with a single hidden layer of 64 units and a scalar output, which produces the final model score used for candidate ranking. 
    }
    \label{fig:WaveletNet_workflow}
\end{figure*}

In this work, we design a network, \texttt{WaveletNet}, to characterize noisy strain data in the vicinity of gravitational-wave candidates. \texttt{WaveletNet} incorporates learnable Morlet wavelets to construct glitch-like templates and performs matched filtering over an extended region surrounding candidate triggers. By explicitly modeling the time--frequency morphology of instrumental glitches, the wavelet-based approach provides a strong inductive bias that is much more sample efficient than generic CNNs when applied to non-local data. We apply our framework to model and characterize glitches in data from the LIGO--Virgo--KAGRA O3 observing run. It is also worth noting that wavelet-based approaches have been successfully applied in other areas of astrophysics and cosmology to achieve robust inference and improved data efficiency compared to generic deep neural networks (e.g., Ref.~\cite{Cheng:2020qbx}).

When integrated modularly into the \texttt{IAS-HM} search pipeline, \texttt{WaveletNet} yields improvements of up to $15\%$ in sensitive spacetime volume, demonstrating its effectiveness as a complementary ranking component rather than a replacement for existing pipelines. This builds on the \texttt{TIER} methodology introduced in Ref.~\cite{Wadekar:2025lhk}. Instrumental glitches often ``come with friends,'' meaning they cluster in time and occur in elevated noise environments\footnote{Note that this behavior has previously been used only in a limited, manual way to reduce contamination in the GWTC-4 catalog (see Fig.~9 of Ref.~\cite{LVK25_GWTC4}).}. Typical current search pipelines evaluate candidate significance using only a narrow ($\sim \pm0.1\,\mathrm{s}$) window around the trigger. As a result, informative structure in the surrounding strain data is ignored. \texttt{WaveletNet} addresses this limitation by producing an external score from nonlocal data that can be modularly combined with the existing ranking statistic.

The paper is organized as follows. In Section~\ref{sec:architecture}, we describe the architecture of \texttt{WaveletNet}. Section~\ref{sec:toy_model} presents tests of our methodology on a simplified toy dataset. In Section~\ref{env_info}, we discuss how extended strain information is incorporated into the search ranking statistic. Section~\ref{training_data} describes the data used to train the ML models and the summary statistics employed to compress the extended strain information. In Section~\ref{sec:results}, we present the resulting sensitivity improvements of the search pipeline, and in Section~\ref{sec:discussion}, we discuss the implications of our methodology.

\subsection{Motivation for an Inductive-Bias-Driven Architecture}

Our methodology is guided by the principle that inference performance can be improved by introducing an explicit inductive bias based on domain knowledge, rather than relying entirely on data-driven feature learning. 
In many signal detection problems in physics, the signals of interest have partially known structure, while the main challenge comes from noise and limited data.

Widely used deep learning approaches (such as CNNs) treat this as a representation learning problem, where the model must learn both the signal structure and the decision boundary from raw data. While flexible, this typically requires large models and datasets, and offers limited physical interpretability. In contrast, classical signal processing provides established tools, such as matched filtering, that are optimal for improving SNRs when both the signal shape and noise distribution is partially or entirely known. Our approach combines these ideas by using flexible signal-processing structure into the matched-filtering step of the pipeline. 

\begin{figure*}
    \centering
    \includegraphics[width=1\linewidth]{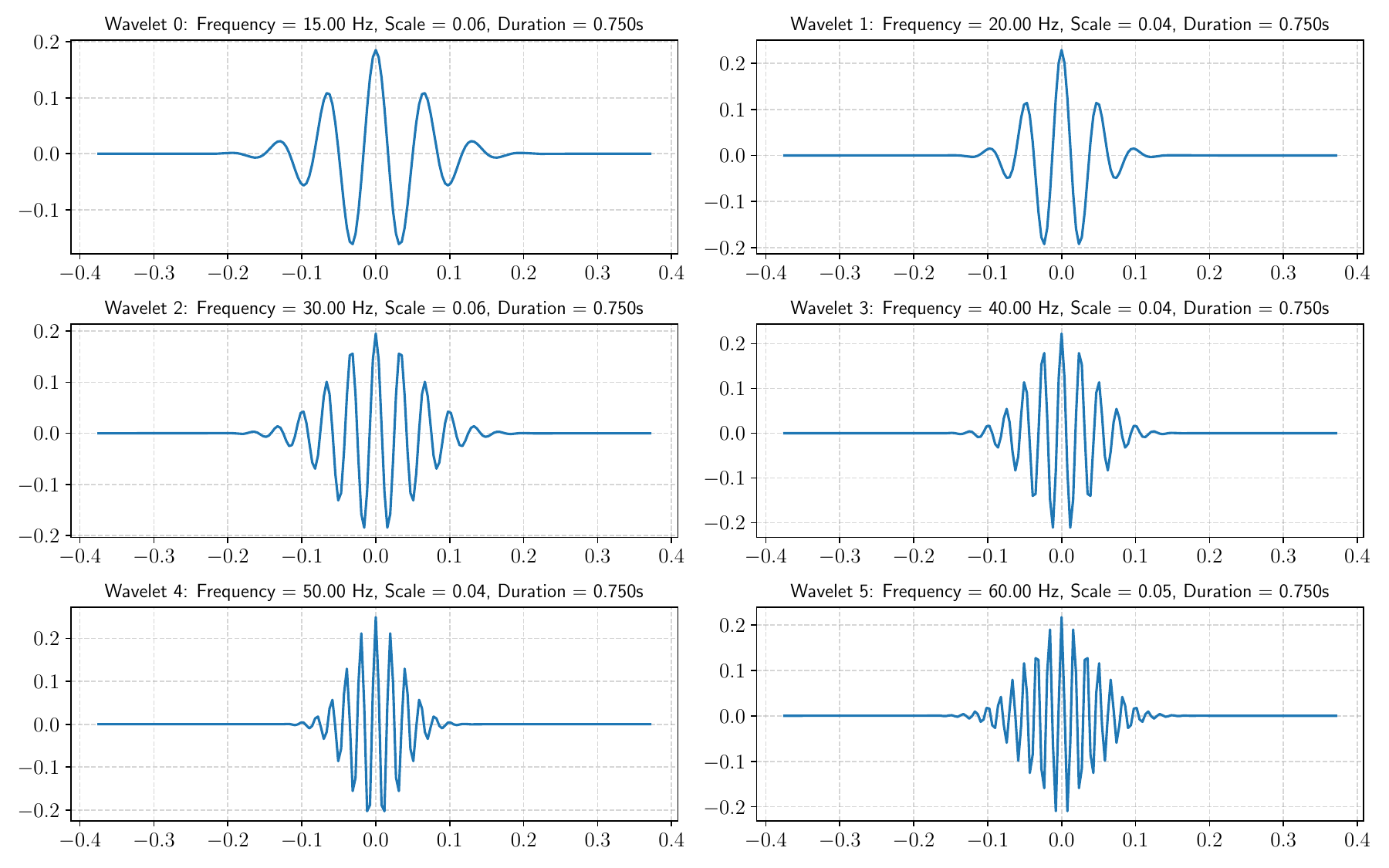}
    \caption{Learned wavelets corresponding to a high-mass template bank (GW signals with total mass $M_\mathrm{tot} \approx 250 M_{\odot}$). Only the scale parameter of each wavelet is trained in our case. The central frequency and duration are held fixed to reduce computational cost (we find that varying them does not yield measurable performance gains in the O3 data). The resulting set of learned waveforms effectively function as a bank of glitch templates for this mass range.
    }
    \label{fig:plots/learned_wavelets_14}
\end{figure*}

\section{WaveletNet Architecture} \label{sec:architecture}
The \texttt{WaveletNet} architecture is broadly divided into two components: a matched filtering module based on learnable Morlet wavelets, and a MLP that combines the resulting features with the summary-statistics input to produce a final candidate score. The workflow of the model is visualized in Fig.~\ref{fig:WaveletNet_workflow}. The following sections describe each component in detail.

\subsection{Learnable Wavelet Matched Filtering Module}

Glitches that resemble high-mass binary black hole signals are empirically observed to possess wavelet-like structure \cite{Hourihane:2022doe,Chatziioannou:2021ezd, Cornish:2025awt, Cornish:2014kda}. Leveraging this property, we employ simple neural networks to learn an optimal wavelet family for modeling glitches using \texttt{WaveletNet}.

The model performs matched filtering on strain data using the learnable Morlet wavelets.
We initialize six Morlet wavelets with center frequencies spanning 15\,Hz to 60\,Hz, covering the frequency band in which most glitches occur. These frequencies are chosen to facilitate stable model training; we find adding additional wavelets does not yield improved performance. The Morlet wavelets are defined as:

\begin{equation}
    \psi(t) = \pi^\mathrm{-1/4} \cdot e^{i2\pi ft} \cdot e^{-t^2/2}\,
\end{equation}
The wavelet has three key components: frequency, scale, and duration.  
The latter two are intrinsically related to each other. In our model, we choose the scale of the wavelet to be the only learnable parameter. The central frequency and duration are held fixed to reduce computational cost, as we find that varying them does not yield measurable performance gains in the O3 data. After scaling, the Morlet wavelet is represented as
\begin{equation}
    \psi_s(t) = \frac{1}{\sqrt{s}} \cdot \psi(\frac{t}{s})\,.
\end{equation}
We trained the scale parameters in logarithmic space to prevent convergence to zero. For example, the trained wavelets are visualized in Fig.~\ref{fig:plots/learned_wavelets_14}. 

The matched filtering of the wavelets and the data was performed using the \texttt{conv1d} function from the \texttt{PyTorch} library. The matched filtering output $y(t)$ is obtained by convolving the input strain data $d_\mathrm{nonlocal}(t)$ with the scaled Morlet wavelet $\psi_s(t)$,
\begin{equation}
    y(t) = (d_\mathrm{nonlocal} * \psi_s^* )(t).
\end{equation}

To maximize the matched-filter response, we apply an analytical phase maximization. This is equivalent to projecting the complex response onto its optimal phase, yielding the magnitude of the output,
\begin{equation}
\tilde{y}(t) = |y(t)| = \sqrt{\operatorname{\Re}(y(t))^2 + \operatorname{\Im}(y(t))^2}\,.
\end{equation}
This operation maximizes the response over the relative phase between the data and the wavelet and is equivalent to an explicit phase rotation that aligns the complex output along the real axis.

Instead of passing the full response to the subsequent MLP, we extract the three highest peaks of $\tilde{y}(t)$ along with their corresponding time indices. The choice to retain only three peaks is motivated by the need to keep the feature dimensionality manageable, and by the empirical observation that our data samples rarely contain more than three prominent glitches. The peaks of $\tilde{y}(t)$ correspond to the amplitudes of the loudest glitch-like responses identified by the wavelet filter, providing a compact summary of the most significant transient activity in the data.

Peaks are identified by segmenting the response into 0.1\,s intervals and selecting the maximum value within each segment. This representation yields improved classification performance while significantly reducing the training time.

\begin{figure*}
    \centering
    \includegraphics[width=0.49\textwidth]{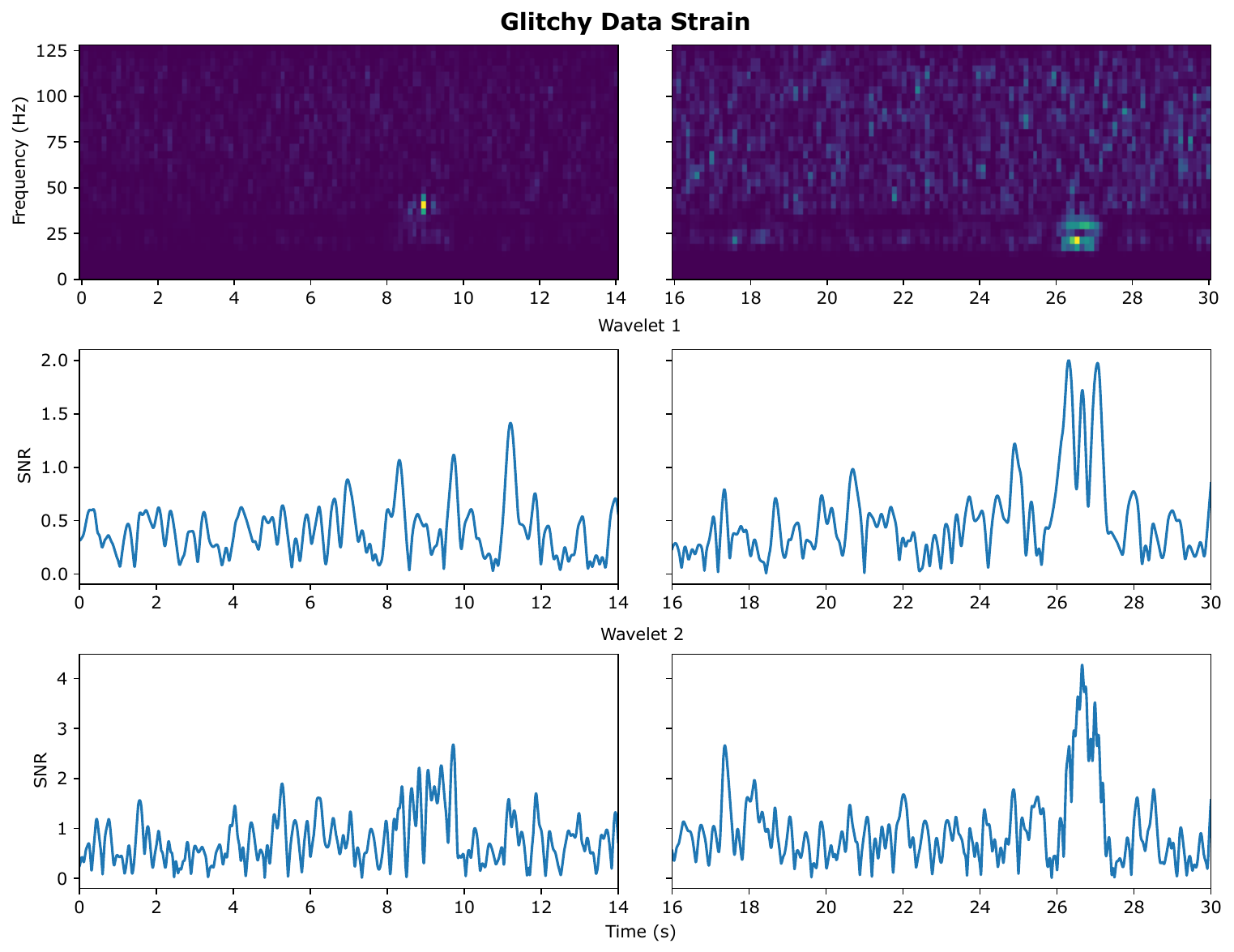}
    \includegraphics[width=0.49\textwidth]{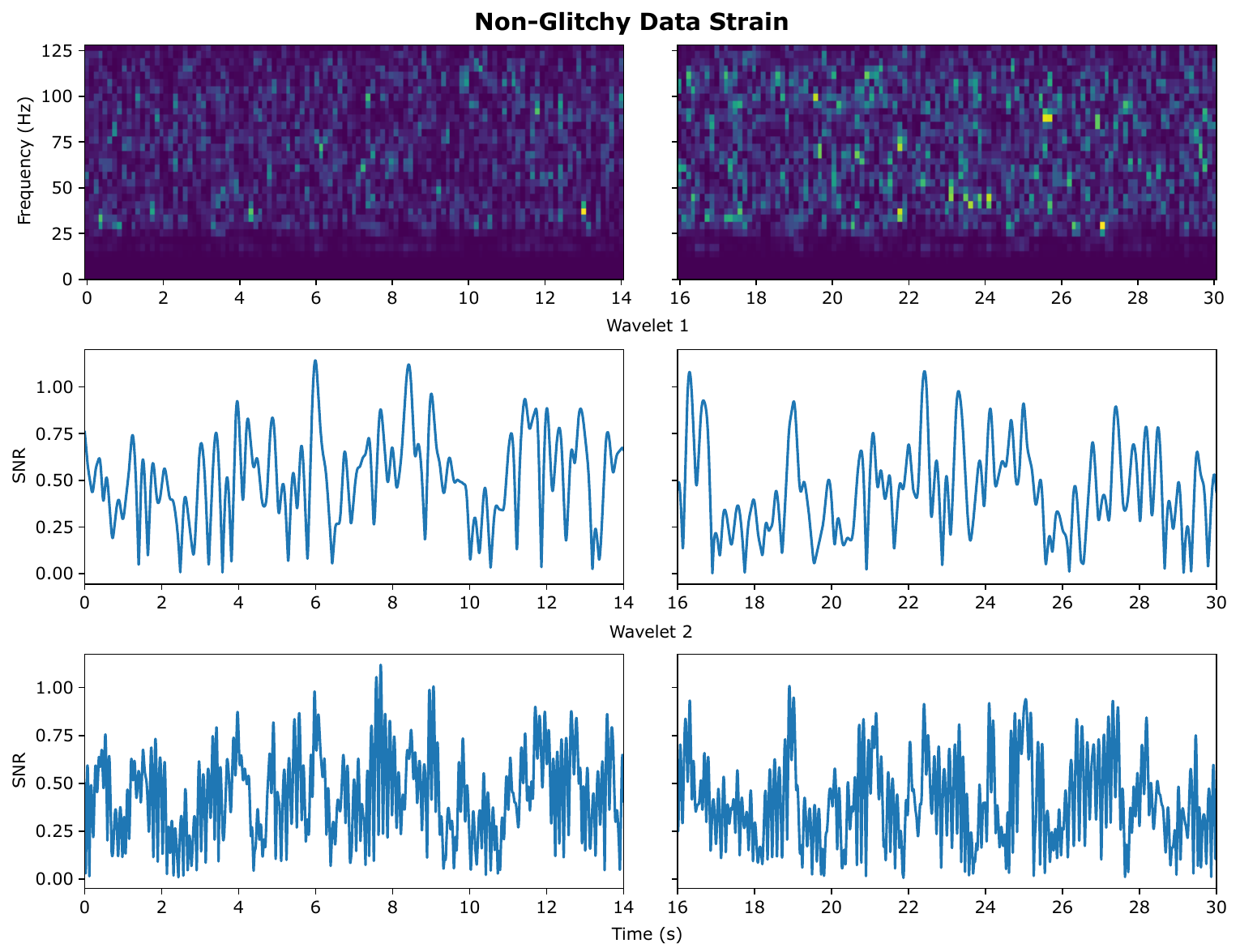}
    \caption{Examples of SNR responses of the wavelet with a glitchy data strain (left panels) and a non-glitchy data strain (right panels). 
    The glitchy data strain containing higher non-Gaussian noise shows high SNR peaks response with wavelets compared to the non-glitchy data strain. Note that the SNR values on the y-axis are different between different panels. }
    \label{fig:result_compare}
\end{figure*}

In Fig.~\ref{fig:result_compare} we compare the matched filtering SNR responses of representative wavelets to a glitchy data strain and to a non-glitchy data strain. For the glitchy strain, the wavelet responses show distinct SNR peaks at times corresponding to non-Gaussian transients that are visible in the spectrograms. In contrast, the responses for the non-glitchy strain remain comparatively low and lack coherent peaks, indicating the absence of significant transient noise. This contrast demonstrates that the learned wavelets are selectively sensitive to glitch-like structure in the nonlocal data.

\subsection{MLP Classification Head}
\label{subsec:mlp_head}

The MLP takes as input the concatenation of wavelet-derived features and nonlocal summary statistics, as discussed in Section~\ref{subsec:nonlocal}. From each wavelet response, the three largest peaks and their corresponding time indices are extracted, yielding a total of 36 wavelet-based features across the full wavelet bank. These features are concatenated with 20 summary statistical features, forming a 56-dimensional input vector. The input is processed by a fully connected layer with Rectified Linear Unit (ReLU) activation, followed by a final linear layer that produces a scalar output score.

Training is performed using the Adam optimizer with a learning rate of $10^{-4}$ and binary cross-entropy loss with logits. The validation loss is monitored to enable early stopping and adaptive learning rate reduction via ReduceLROnPlateau schedule, and the model state corresponding to the lowest validation loss is retained.

In this work, the ML component is designed to augment the traditional matched filtering search pipeline, rather than replace it. The glitches used for training are noise transients that produce triggers in the existing GW template banks and therefore span a range of morphologies, particularly in their characteristic frequency content. Capturing this diversity with a single wavelet model is challenging. We thus train separate instances of the models for triggers from different GW template banks (the banks are approximately partitioned by total mass, see \cite{Wad23_TemplateBanks} for details). The different banks correspond to waveforms with characteristic GW signal frequency, hence glitches associated with a given bank tend to exhibit similar time-frequency structure. Therefore, for each bank, the model is trained exclusively on glitches identified by matched filtering with templates from that bank. During inference, the probability of a given candidate trigger being a glitch is computed using the model corresponding to the bank of the triggering template.

\section{Performance on a toy dataset}
\label{sec:toy_model}

To compare the performance of CNNs with our \texttt{WaveletNet} model, we perform a toy experiment comparing a generic CNN with a simpler version of our wavelet-based model from the above section~\ref{sec:architecture}. The dataset used in this section is a synthetic time series dataset designed to study binary signal detection under controlled conditions. Each sample consists of a 10 s, one-dimensional time series sampled at 256 Hz. There are two classes of training samples, the first class contains pure Gaussian noise, while the second class contains a weak, localized signal embedded in Gaussian noise. 

The signal is a Morlet wavelet with fixed time–frequency structure, randomly positioned within the time series for temporal invariance. The signal amplitude is chosen such that the resulting SNR is moderate, rendering the signal typically indistinguishable by eye. All datasets are class-balanced and generated deterministically, with multiple dataset sizes considered to assess scaling of ML models with available training data.

We first apply matched filtering using learnable wavelets to emphasize relevant time-frequency structure in the data. From the resulting responses, we extract a small set of summary statistics, such as the mean, standard deviation, range, and peak prominence, to characterize signal strength and distinctiveness. This structured representation reduces model complexity, stabilizes training, and preserves interpretability while retaining the benefits of ML-based optimization.

\begin{figure}
    \centering
    \includegraphics[width=1\linewidth]{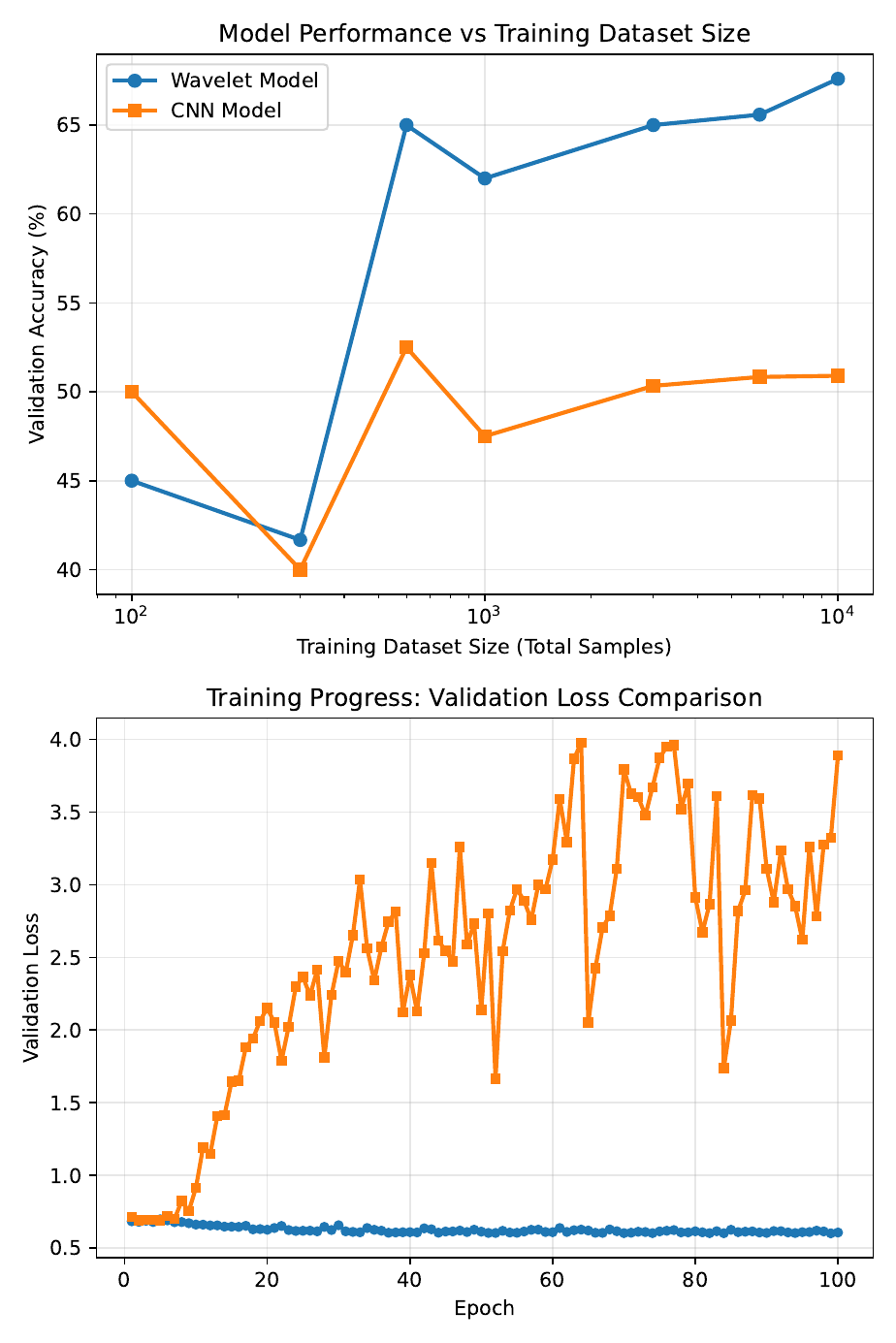}
    \caption{Comparison of a wavelet-based neural network and a CNN on a toy-model dataset (see Section~\ref{sec:toy_model} for details). This figure illustrates the effect of the inductive bias used in \texttt{WaveletNet} on data efficiency and training stability. Top panel: Validation accuracy as a function of training dataset size for a wavelet-based model and a CNN, showing that the wavelet-based model achieves higher accuracy with fewer training samples. Bottom panel: Validation loss as a function of training epoch for the same models, obtained using a fixed training set of $400$ examples per class. The wavelet-based model exhibits stable convergence, while the CNN shows significantly larger fluctuations. These results qualitatively demonstrate the advantage of incorporating signal-processing structure in limited-data regimes.
}
    \label{fig:toy_model_plots}
\end{figure}

In Fig.~\ref{fig:toy_model_plots} we show the validation performance of the two models as a function of training dataset size and training epoch. The wavelet-based model achieves higher validation accuracy with significantly fewer training samples and exhibits stable convergence during training. In contrast, the CNN shows slower improvement with dataset size and substantial instability in validation loss, reflecting its higher data requirements and sensitivity to overfitting. These trends qualitatively demonstrate how incorporating problem-specific structure can improve data efficiency and training robustness, even in simplified settings. Note that for very low training dataset size, there can be significant stochasticity in training the models, this is likely the reason for change in relative behavior for the in the left panel of Fig.~\ref{fig:toy_model_plots}. 

More broadly, synergies between wavelet-based representations and flexible deep learning architectures have been explored in the machine learning literature, demonstrating that combining explicit time--frequency structure with learnable models can improve robustness and generalization in limited-data regimes~\cite{Oyallon_2019,Michau_2022,Khemani2022,bruna2012invariantscatteringconvolutionnetworks}.

\section{Including environmental information in the search ranking statistic}
\label{env_info}

When searching for GW signals, each search pipeline finds a number of signal candidates.
To determine the significance of the candidates, a ranking score/statistic is assigned to each candidate. The optimal ranking score is given by the Neyman-Pearson lemma, and corresponds to the ratio of the likelihood of the candidate being either a signal ($\mathcal{S}$) or noise ($\mathcal{N}$):
\begin{equation} 
    \mathcal{L} \equiv \frac{P(d|\mathcal{S})}{P(d|\mathcal{N})}
    \equiv \frac{P(d_\mathrm{local}, d_\mathrm{nonlocal}|\mathcal{S})}
    {P(d_\mathrm{local}, d_\mathrm{nonlocal}|\mathcal{N})}\,, \label{eq:NeymanPearson}
    \end{equation}
where $d_\mathrm{local}$ is the region within ±0.1 s of the candidate, and $d_\mathrm{nonlocal}$ extends to ±15 s of the candidate, as will be elaborated in the following sections. Expanding Eq.~\eqref{eq:NeymanPearson}, we obtain:
\begin{equation}\label{eq:NeymanPearson1}
 \mathcal{L} \equiv  \frac{P(d_\mathrm{local}|\mathcal{S})}{P(d_\mathrm{local}|\mathcal{N})}
    \frac{P(d_\mathrm{nonlocal}|d_\mathrm{local},\mathcal{S})}
    {P(d_\mathrm{nonlocal}|d_\mathrm{local},\mathcal{N})}\,.
\end{equation}

Calculating $P(d_\mathrm{nonlocal}|d_\mathrm{local},\mathcal{N})$ is challenging because of its non-Gaussian and nonstationary nature. Therefore, search pipelines often use the approximation 
\begin{equation} \label{eq:pipeline}
    \mathcal{L}_{\rm pipeline} \approx \frac{P(d_{\rm local} | \mathcal{S})}{P(d_{\rm local} | \mathcal{N})}\,,
\end{equation}
which effectively assumes the second fraction to be unity. This corresponds to treating the noise far from the candidate event as statistically independent of the local data $d_\mathrm{local}$.

However, as mentioned above, glitches can ``come with friends'': if $d_\mathrm{local}$ contains a glitch (i.e., the true hypothesis is $\mathcal{N}$), the surrounding data $d_\mathrm{nonlocal}$ is often contaminated by elevated noise levels. Consequently, the standard approximation is not strictly valid, since the statistical properties of $d_\mathrm{nonlocal}$ need not resemble those of a typical Gaussian background.

\begin{figure}
    \centering
    \includegraphics[scale=0.4,keepaspectratio=true]{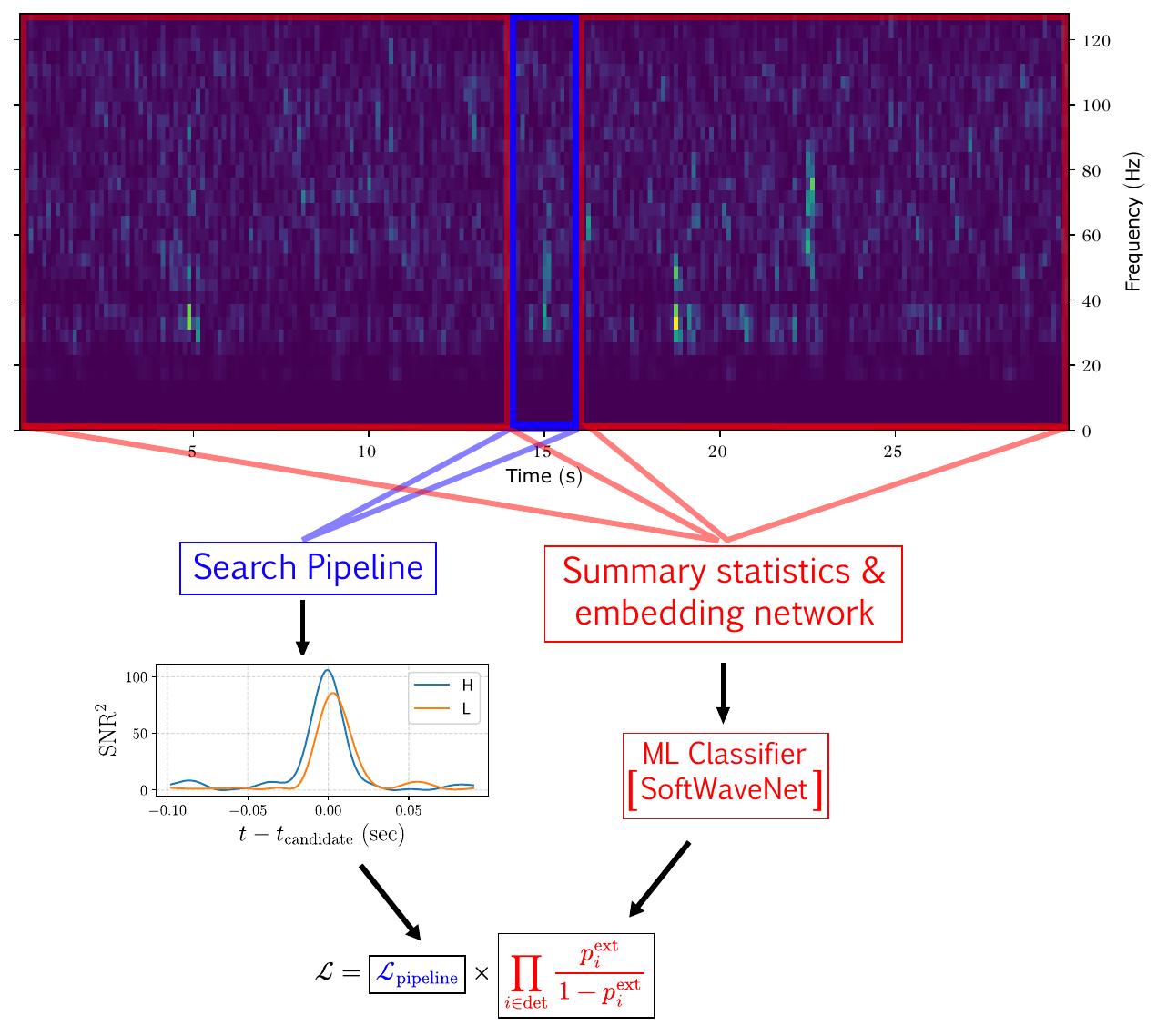}
    \caption{The ranking statistic of a typical search pipeline ($\mathcal{L}_\mathrm{pipeline}$) uses local SNR from matched filtering to sort or rank a candidate event by its likelihood of being an astrophysical signal rather than noise. The \texttt{TIER} framework can be used to augment the ranking statistic of a pipeline by including complementary information about the extended strain data next to the candidate time (see Eq.~\eqref{eq:NeymanPearson}). We thus obtain the probability that the candidate is an astrophysical signal ($p^\mathrm{ext}$) based on the extended strain data and include it in the candidate ranking statistic.} 
    \label{fig:TIER}
\end{figure}

We use our ML method to address this issue; we learn the distribution of non-Gaussian noise from the detector data and down-weight non-Gaussian transients. As shown schematically in Fig.~\ref{fig:TIER}, the resulting ML score can be modularly combined with the traditional pipeline statistic to further calibrate the ranking score:
\begin{equation}
        \mathcal{L}_\mathrm{new} \simeq \mathcal{L}_\mathrm{pipeline} \times \prod_{i\in \mathrm{det}} \frac{p^\mathrm{ext}_i}{1-p^\mathrm{ext}_i}\,.
        \label{eq:NeymanPearson3}
\end{equation}

Here $p^\mathrm{ext}_i$ is the classifier output trained to estimate 
$p^\mathrm{ext}_i = p(d|\mathcal{S})/[p(d|\mathcal{S})+p(d|\mathcal{N})]$ in detector $i$. The ratio $p^\mathrm{ext}_i/(1-p^\mathrm{ext}_i)$ therefore yields $p(d|\mathcal{S})/p(d|\mathcal{N})$, providing a direct estimate of the likelihood ratio contribution from the nonlocal data that can be consistently combined with $\mathcal{L}_\mathrm{pipeline}$.
The new likelihood $\mathcal{L}_\mathrm{new}$ should approximate the true likelihood better than $\mathcal{L}_{\rm pipeline}$, as calculated in Eq.~\eqref{eq:pipeline}. 
  
\section{Training Data} \label{training_data}
In this section, we describe the input data streams used for training \texttt{WaveletNet} and obtaining $p^\mathrm{ext}_i$. \texttt{WaveletNet} integrates two inputs: a whitened strain time series from the nonlocal region surrounding each candidate, and a vector of statistical features summarizing the candidate’s properties. Together, these inputs provide complementary contextual and summary information. The following sections describe these inputs in detail. Separate models are trained for each of the 17 binary black hole template banks, whose construction and bank boundaries are described in Section~II of~\cite{Wadekar:2024zdq}. For each bank, the models are trained using approximately $15{,}000$ injection examples and $23{,}000$ glitch examples.

\subsection{Whitened Time Series}
We extract time-series data segments of length 30 s from the nonlocal region surrounding each candidate event. These segments are selected to exclude the immediate vicinity of the trigger, ensuring that the model captures background noise characteristics rather than features of the candidate itself. By providing raw contextual information about the surrounding noise environment, the time-series input enables \texttt{WaveletNet}'s learnable wavelet filters to characterize noise properties that are relevant for classification.

\subsection{Additional nonlocal statistical features of the GW candidates}\label{subsec:nonlocal}
In addition to the time series input described above, we incorporate a set of summary statistical features derived from the nonlocal data surrounding each candidate. These features, provided by the \texttt{IAS-HM} search pipeline, offer a compact description of the broader data environment and are intended to complement the information extracted directly from the strain time series.

We present below the statistical features computed from the nonlocal region of the data. These features were developed through empirical, trial and error exploration, with the aim of capturing characteristics of the surrounding data that may indicate elevated noise levels or nonstationarity. Similar summary statistics are also used in the random forest models described in Ref.~\cite{Wadekar:2025lhk}. We note that the development of more informative summary statistics, or the use of ML based compression methods for the environmental data, may further improve our ML classifier performance.

\begin{enumerate}
    \item $\Delta t_\mathrm{triggers}$: This quantity denotes the time interval between the candidate and the other neighboring triggers, in the range of 0.5\,s $<|t-t_\mathrm{candidate}| < 15$\,s, in each detector's catalog picked up in the matched filtering stage of the pipeline. We only include binary black holes with individual masses $>10 M_\odot$, which safely have duration shorter than 0.5\,s, to avoid signal overlap.
    \item SNR of nearby triggers: We include information about the SNRs of neighboring triggers observed in each detector. We use full-harmonic SNRs calculated as $|\rho^2_\mathrm{HM}|\equiv|\rho_{22}|^2+|\rho^\perp_{33}|^2+|\rho^\perp_{44}|^2$, where $\rho^\perp_{33}$ and $\rho^\perp_{44}$ correspond to the SNR of the orthogonalized higher-harmonic templates. We rescale these values to reduce the numerical range of the inputs for the ML models. 
    \item Template ID of the neighboring triggers: In the IAS pipeline, the template ID corresponds to singular vector components ($c_0, c_1$)~\cite{Roulet:2019hzy, Wadekar:2023kym}. Glitches are caught more frequently by certain templates (short-duration waveforms) and providing this information to the ML model could be useful. 
    \item $\int\frac{|h_\mathrm{norm}(f)|^2}{\mathrm{PSD}_\mathrm{local}(f)}df$: This quantity measures the sensitivity of a given local strain data segment to gravitational waveforms from a particular template bank. Essentially, this quantity penalizes data segments with high PSD values, often dominant in the segments with gravitational strains. 
    \item $\Delta t_\mathrm{hole}$: Distance to the nearest ``hole''. These are the regions excised by the IAS-HM search pipeline. They correspond to very loud noise transients, and are expected to have SNR $\gtrsim$ 20. We include this parameter separately from the nearby triggers, as these data regions do not have an accurate SNR estimate in our pipeline (as we inpaint such loud noise data segments before estimating the PSD). 
    \item $N_\mathrm{noisy\, bands}$: These are the number of noisy frequency bands/channels overlapping with the triggers. We use results from the \texttt{Band Eraser} tool from Ref.~\cite{Wadekar:2024zdq} to estimate this quantity. The \texttt{Band Eraser} tool removes individual noisy segments with dimensions $\rm 64\,s \times 2\,Hz$ in the spectrogram of the strain data.
\end{enumerate}

\begin{figure*}
    \centering
    \includegraphics[width=1\linewidth]{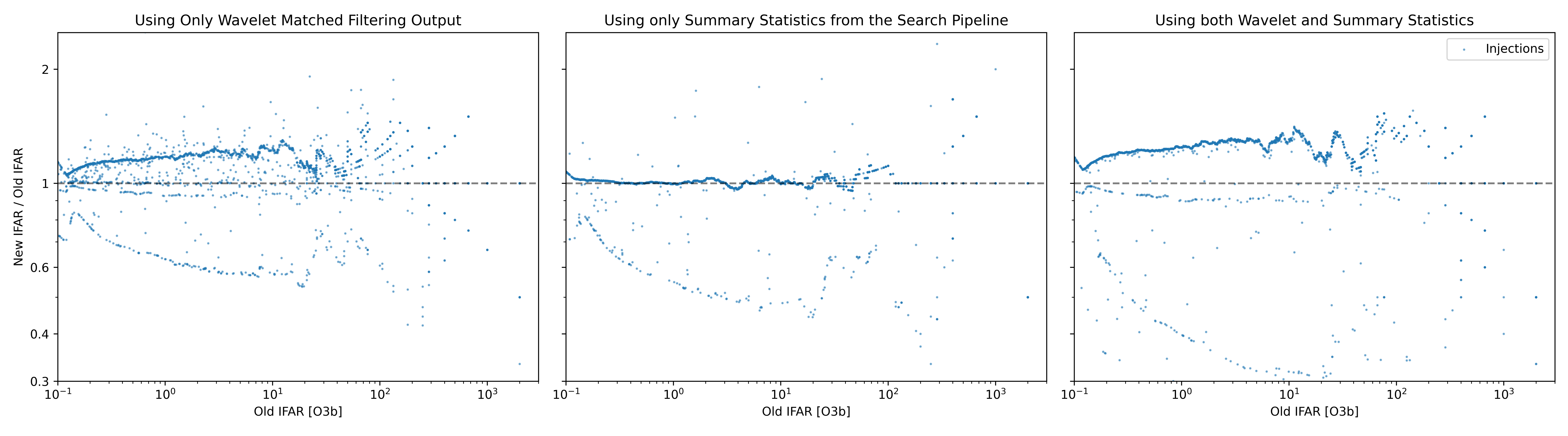}
    \caption{The two classes of inputs to \texttt{WaveletNet} are the SNR responses obtained from matched filtering with the learned wavelets, and the additional summary statistics provided by the search pipeline (see Fig.~\ref{fig:WaveletNet_workflow}). Here, we illustrate the relative importance of these two input classes by quantifying their effect on the improvement in IFAR for injections in the data. The left panel shows the performance obtained using only the wavelet matched filtering outputs. The center panel shows the IFAR improvement obtained when using only the summary statistics from the search pipeline (see section~\ref{subsec:nonlocal}). The right panel shows the results for the full \texttt{WaveletNet} model, which combines both the wavelet-based features and the pipeline summary statistics. We find that the wavelet-based features provide a stronger contribution to the IFAR improvement than the pipeline summary statistics alone.\label{fig:IFAR_compare}
    }
\end{figure*}

\section{Search pipeline Sensitivity Improvement}\label{sec:results}


We apply the trained \texttt{WaveletNet} models to candidates produced by the \texttt{IAS-HM} search pipeline, computing the external probability $p^\mathrm{ext}$ for both background and coincident events. These values are combined with the traditional ranking statistic using Eq.~\eqref{eq:NeymanPearson3} to form a new ranking statistic. This re-ranking modifies the false alarm rate (FAR) and the inferred astrophysical probability, $p_\mathrm{astro}$, assigned to each candidate.

As a first diagnostic of the trained models, we examine the learned wavelet filters.
In Fig.~\ref{fig:plots/learned_wavelets_14} we show the trained wavelets for one representative template bank, BBH-14, corresponding to total masses of $M_\mathrm{tot} \approx 250\,M_{\odot}$. These learned wavelets can be interpreted as data-driven templates for glitch-like transients present in the environmental channels.

To translate this wavelet-level sensitivity into an impact on search results, we compute an external probability score, $p^{\rm ext}$, for each coincident candidate identified by the \texttt{IAS-HM} pipeline. For a given candidate, the learned wavelet responses and nonlocal summary statistics are processed by \texttt{WaveletNet} to produce $p^{\rm ext}_i$ for each participating detector. These scores are then combined with the traditional pipeline ranking statistic $\mathcal{L}_{\rm pipeline}$ according to Eq.~\eqref{eq:NeymanPearson3} to form a re-ranked statistic $\mathcal{L}_{\rm new}$, yielding an updated ordering of candidates.

In Fig.~\ref{fig:IFAR_compare} we illustrate the resulting changes in inverse false-alarm rate (IFAR) for candidates in the BBH-14 template bank obtained using this method. Here, the old IFAR refers to the original ranking produced by the \texttt{IAS-HM} pipeline without \texttt{WaveletNet}, and we plot the ratio of the updated IFAR to the original value after incorporating $p^{\rm ext}$. For many candidates, particularly those with modest IFAR in the original ranking, the inclusion of $p^{\rm ext}$ leads to a noticeable suppression of events occurring in noisy nonlocal environments, while leaving cleaner candidates largely unaffected or modestly enhanced. This behavior is consistent with the intended role of $p^{\rm ext}$ as a noise-aware correction to the pipeline ranking statistic.

Using these re-ranked candidate lists, we then compute updated false-alarm rates and astrophysical probabilities for simulated signal injections. The sensitive spacetime volume, $VT$, quantifies the four-dimensional volume over which a search is sensitive to astrophysical signals and is the primary figure of merit for comparing detection efficiency at fixed false-alarm rate; improvements in $VT$ correspond directly to an increased expected detection rate. In its simplest form, $VT$ is often estimated as

\begin{equation}
VT \approx \frac{N_\mathrm{det}}{N_\mathrm{inj}} \, V_\mathrm{inj} T_\mathrm{inj}\,,
\end{equation}
where $N_\mathrm{det}$ is the number of detected injections above a given threshold, $N_\mathrm{inj}$ is the total number of injections, and $V_\mathrm{inj} T_\mathrm{inj}$ is the spacetime volume over which injections are distributed.

Our computation of the volume-time, $\overline{VT}$, employs a more accurate weighted Monte Carlo integral over the injections recovered by the pipeline with $\mathrm{IFAR} > 1$~yr, following~\cite{Tiw17_VT_estimation}:
\begin{equation}
\overline{VT} \simeq \frac{T_{\mathrm{obs}}}{N_{\mathrm{draw}}}\sum_{j=1}^{N_{\mathrm{found}}}
\frac{f(z_j)\,\dfrac{1}{1+z_j}\,\dfrac{dV_c(z_j)}{dz}\,p(\theta_j)}{\pi_{\mathrm{draw}}(\theta_j, z_j)}\,,
\label{eq:VT_est}
\end{equation}
where $T_{\mathrm{obs}}$ is the observation time of the analyzed dataset, $p(\theta)$ is the astrophysical probability distribution over the intrinsic source parameters $\theta = (m_1^{\mathrm{s}}, m_2^{\mathrm{s}}, \vec{s}_1, \vec{s}_2)$, $V_c$ is the comoving volume, $z$ is the redshift, $N_{\mathrm{draw}}$ is the total number of injections generated from the sampling distribution, and $\pi_{\mathrm{draw}}(\theta, z)$ is the corresponding injection sampling probability elaborated in Ref.~\cite{Mehta:2025jiq}. 

\begin{figure}
    \centering
    \includegraphics[width=1\linewidth]{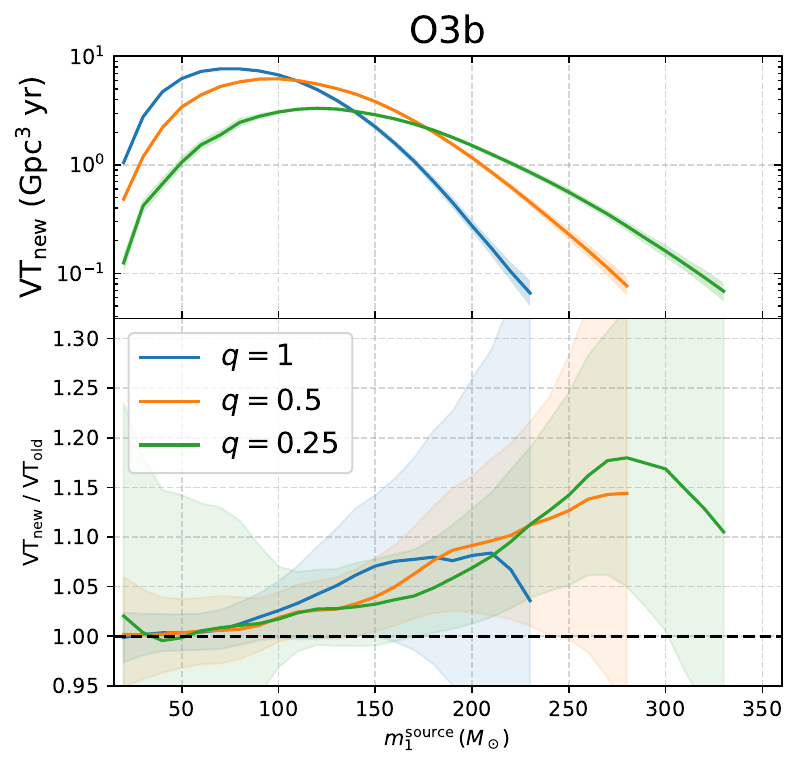}
    \caption{Improvement in the sensitive spacetime volume ($VT$) of the \texttt{IAS-HM} search obtained by incorporating \texttt{WaveletNet}. The curves labeled ``old'' correspond to the $VT$ reported in Ref.~\cite{Mehta:2025jiq}, which does not include \texttt{WaveletNet}. We use the injection catalog released by the LVK Collaboration on \texttt{Zenodo}~\cite{zenodoLVK}. The shaded bands denote the $90\%$ confidence intervals obtained via bootstrap resampling of the injection set. The observed sensitivity gains are primarily concentrated at high total masses and asymmetric mass ratios, corresponding to short-duration signals that are most susceptible to contamination from transient noise. We also observe an increase in the significance of several near-threshold GW candidates identified by the \texttt{IAS-HM} search.}
    \label{fig:VT_increase}
\end{figure}

In Fig.~\ref{fig:VT_increase} we show the resulting improvement in $VT$ as a function of inverse false-alarm rate (IFAR) when incorporating $p^\mathrm{ext}$ into the ranking statistic. The shaded bands denote the $90\%$ confidence intervals obtained via bootstrap resampling of the injection set. Across a wide range of IFAR thresholds, we observe a systematic increase in $VT$, indicating enhanced detection efficiency at fixed false-alarm rate.

\section{Discussion}
\label{sec:discussion}

The primary goal of this work is to introduce and evaluate a wavelet-based approach for characterizing nonlocal noise environments surrounding GW candidates, and to assess its impact when integrated into existing ranking pipelines. In this section, we discuss how \texttt{WaveletNet} compares to \texttt{TIER}, introduced in Ref.~\cite{Wadekar:2025lhk}, and highlight the regimes in which the wavelet-based framework provides complementary or potentially improved performance.

A key distinction between \texttt{WaveletNet} and \texttt{TIER} lies in how nonlocal noise is modeled. The \texttt{TIER} method relies on matched filtering with template banks (originally constructed to identify compact binary coalescence signals). While this approach is effective at capturing certain forms of excess power, these templates are not optimized for instrumental glitches, which often exhibit time-frequency morphologies that can differ substantially from those of astrophysical signals. In contrast, \texttt{WaveletNet} learns a set of data-driven wavelet templates directly from environmental data, enabling it to adapt to the characteristic time-frequency structure of glitches. This design choice allows the model to assess the noise environment surrounding a candidate in a manner more closely aligned with the underlying glitch phenomenology.

Specifically, \texttt{TIER} uses the same template banks employed by the \texttt{IAS-HM} pipeline to generate matched filtering triggers for glitch identification. This strategy is computationally efficient, but it represents noise transients solely through their projections onto astrophysical GW templates. As a result, it can be suboptimal when instrumental glitches exhibit morphologies that differ significantly from compact binary coalescence signals.

In terms of detection efficiency, we find that the improvement in sensitive spacetime volume $VT$ achieved on O3 data is comparable to that obtained with the original \texttt{TIER} approach. While the gains are not substantially larger in this dataset, the wavelet-based method is complementary rather than redundant. Because \texttt{WaveletNet} learns glitch morphologies directly from the data, it is naturally suited to adapting to changes in detector behavior. We therefore expect this approach to be particularly valuable in future observing runs, such as O4 and beyond, where the population and morphology of glitches may differ from those seen in O3. As long as glitches admit a reasonably sparse wavelet representation, the method can retrain and recalibrate without relying on fixed signal-based template banks. The wavelet parameters corresponding to glitches which are learnt by \texttt{WaveletNet} can be used as priors in parameter estimation and data cleaning methods such as BayesWave \cite{Cornish:2014kda}.

Unlike approaches that characterize nonlocal regions solely through summary statistics, \texttt{WaveletNet} performs matched filtering against learned wavelet templates, providing a direct and interpretable measure of transient noise activity. This inductive bias leads to improved sample efficiency relative to more generic CNNs, which typically require substantially larger training datasets to learn comparable representations. Moreover, the scalar output score produced by \texttt{WaveletNet} is easily interpretable and can be seamlessly incorporated into existing detection and ranking pipelines.

We also explored several alternative ML approaches based on CNNs, including architectures augmented with attention mechanisms. These models offer greater flexibility and, in principle, can learn a broader class of time--frequency patterns than wavelet-based models. However, in practice, we find that CNN-based approaches require significantly more training data to achieve stable performance and are more prone to overfitting in the limited-data regime relevant for environmental glitch characterization. For a fair comparison, the CNNs were trained using the same dataset sizes as \texttt{WaveletNet}, namely $15{,}000$ injection examples and $23{,}000$ glitch examples per bank. While CNNs may ultimately outperform wavelet-based methods given sufficiently large and diverse training sets, \texttt{WaveletNet} provides a more data-efficient solution for the problem considered here. A detailed discussion of these alternative models and their performance is provided in Appendix~\ref{sec:other_approaches}.

In Appendix~\ref{sec:more_figures}, we compare \texttt{WaveletNet} with the random forest regression model introduced in Ref.~\cite{Wadekar:2025lhk}, as well as with variants of \texttt{WaveletNet} that differ in architectural details. As described in Section~\ref{training_data}, our model integrates two input data streams: wavelet-based features derived from nonlocal strain data and a set of summary statistical features. The additional comparisons presented in Appendix~\ref{sec:more_figures} 
and Fig.~\ref{fig:IFAR_compare} demonstrate that the wavelet-based features play a dominant role in the observed performance improvements, while the summary statistics provide complementary contextual information. This confirms that explicitly modeling glitch morphology via learned wavelets is a key contributor to the effectiveness of the proposed approach.

\section{Conclusions}\label{sec:conclusions}

A key takeaway of this work is that effective nonlocal noise characterization does not require highly complex model architectures. Instead of relying on large, data-hungry CNNs, we introduce a lightweight wavelet-based model that incorporates prior knowledge about glitch structure and can be trained with substantially fewer examples. This simpler architecture is easier to train, more stable in limited-data regimes, and still captures the features needed to improve candidate ranking.

Gravitational-wave search pipelines typically assess the origin of a candidate using data localized to the immediate vicinity of the event. However, information contained in the broader, nonlocal data environment can provide valuable additional context for distinguishing astrophysical signals from instrumental noise. In this work, we develop machine-learning models that analyze the nonlocal environment surrounding a candidate by matched filtering learnable wavelets against the strain data, complemented by sparse summary statistical features.

The scalar output of these models can be straightforwardly incorporated into existing ranking statistics used by search pipelines, enabling re-ranking of candidates without disrupting established detection workflows. Applying this approach to the \texttt{IAS-HM} search, we observe improvements of up to $\sim 15\%$ in sensitive spacetime volume, as shown in Fig.~\ref{fig:VT_increase}, demonstrating enhanced detection efficiency at fixed false-alarm rate.

Importantly, the proposed framework is general and can be integrated with any gravitational-wave search pipeline that produces candidate triggers and associated environmental data. Because the wavelet templates are learned directly from the data, the method is well suited to adapting to changes in detector noise characteristics. Future work will extend this analysis to O4 and subsequent observing runs, where evolving glitch populations and increased detector sensitivity are expected to further benefit from data-driven, adaptive noise characterization.

\section*{Acknowledgments}
A.P., D.W., M.H.Y.C. and E.B.~are supported by NSF Grants No.~AST-2307146, No.~PHY-2513337, No.~PHY-090003, and No.~PHY-20043, by NASA Grant No.~21-ATP21-0010, by John Templeton Foundation Grant No.~62840, by the Simons Foundation [MPS-SIP-00001698, E.B.], by the Simons Foundation International [SFI-MPS-BH-00012593-02], and by Italian Ministry of Foreign Affairs and International Cooperation Grant No.~PGR01167. M.H.Y.C. is a Croucher Fellow supported by the Croucher Foundation. M.H.Y.C. is supported by the Jonathan M. Nelson Center for Collaborative Research at the Institute for Advanced Study. This work was carried out at the Advanced Research Computing at Hopkins (ARCH) core facility (\url{https://www.arch.jhu.edu/}), which is supported by the NSF Grant No.~OAC-1920103.

\appendix

\section{Other approaches}
\label{sec:other_approaches}

In this section, we present alternate ML model architectures that we tried, but that did not yield usable results. While the reason for the failures are often complicated, we present some plausible explanations.

\subsection{Convolutional Neural Network (CNN)}
We experimented with a brute force method of directly training CNNs on the GW strain data examples mentioned in section~\ref{training_data}. We also tried including an attention transformer layer to focus on the glitches. We elaborate the same in the following sections.

\subsubsection{Training on spectrograms of the nonlocal region}\label{specgrams_training}
We generated spectrograms from the time-series data using the \texttt{specgram} function from the \texttt{matplotlib} library. The spectrograms were computed with a sampling frequency \texttt{Fs = 256}, an FFT window size \texttt{NFFT = 64}, and an overlap of \texttt{noverlap = 32} between successive windows. The spectrogram was divided into the left and right region of the candidate and passed separately to the CNN (see Fig.~\ref{fig:TIER}). 

One of the models we explored was a CNN composed of multiple stacked layers of 2D convolutions, batch normalization, ReLU activations, and max pooling. The motive was to learn the spectral patterns found with glitches by learning features in the input spectrograms. We found that the CNN struggled to distinguish between classes effectively.  

We also experimented with a model that incorporated a SpectralAttention  mechanism following several convolutional blocks, inspired by Ref.~\cite{Melchior_2023}. Each convolution block consisted of a 2D convolution layer with progressively larger kernel sizes (5×5), (11×11), (21×21), followed by batch normalization, LeakyReLU activation, and max pooling. After reshaping the output, we applied a custom attention layer that projected the features into query and value representations and used soft attention weights to emphasize salient regions. The intuition was that this mechanism could help the model identify and focus on glitch-specific patterns, especially in noisy or ambiguous cases. However, the attention-enhanced model did not yield a significant improvement. 
A possible explanation is that the spectrograms from the glitch and injection classes often appeared quite similar. This suggests that the CNN may have failed to isolate features that were truly distinctive of glitchy backgrounds, potentially because of subtle underlying spectral differences. 

\texttt{WaveletNet} performs better in this regime despite the limited size of the training set because its wavelet filters act as convolution kernels explicitly optimized to detect non-Gaussian transient structure. By imposing an inductive bias tailored to the morphology of glitches, the model focuses on a restricted and physically motivated feature space rather than learning arbitrary spectral patterns. As a result, \texttt{WaveletNet} achieves improved data efficiency and can be trained effectively with substantially fewer samples than more flexible CNN-based architectures.

\subsubsection{Training on raw Time series}

Another approach we tried included training of one-dimensional CNNs on raw time series without any preprocessing. The motive of this approach was to include the raw amplitude and phase information without any frequency-time trade off and eliminating the dependency on preprocessing parameters. 
Additionally, we experimented with two types of attention mechanisms within our model structure - TransformerSelfAttention and SpectralAttention. 

CNNs can capture short-range temporal features without any complex layers, but to recognize long-term features, they require deep layers. TransformerSelfAttention allows an interconnection to develop between every element of the time series, but unlike RNNs, these interconnections are direct. This allows TransformerSelfAttention to be a global pattern detector, which should have helped in creating a connection between noisy area and glitches. 

\section{More figures and comparisons}
\label{sec:more_figures}

\begin{figure}[t]
    \centering
    \includegraphics[width=0.9\linewidth]{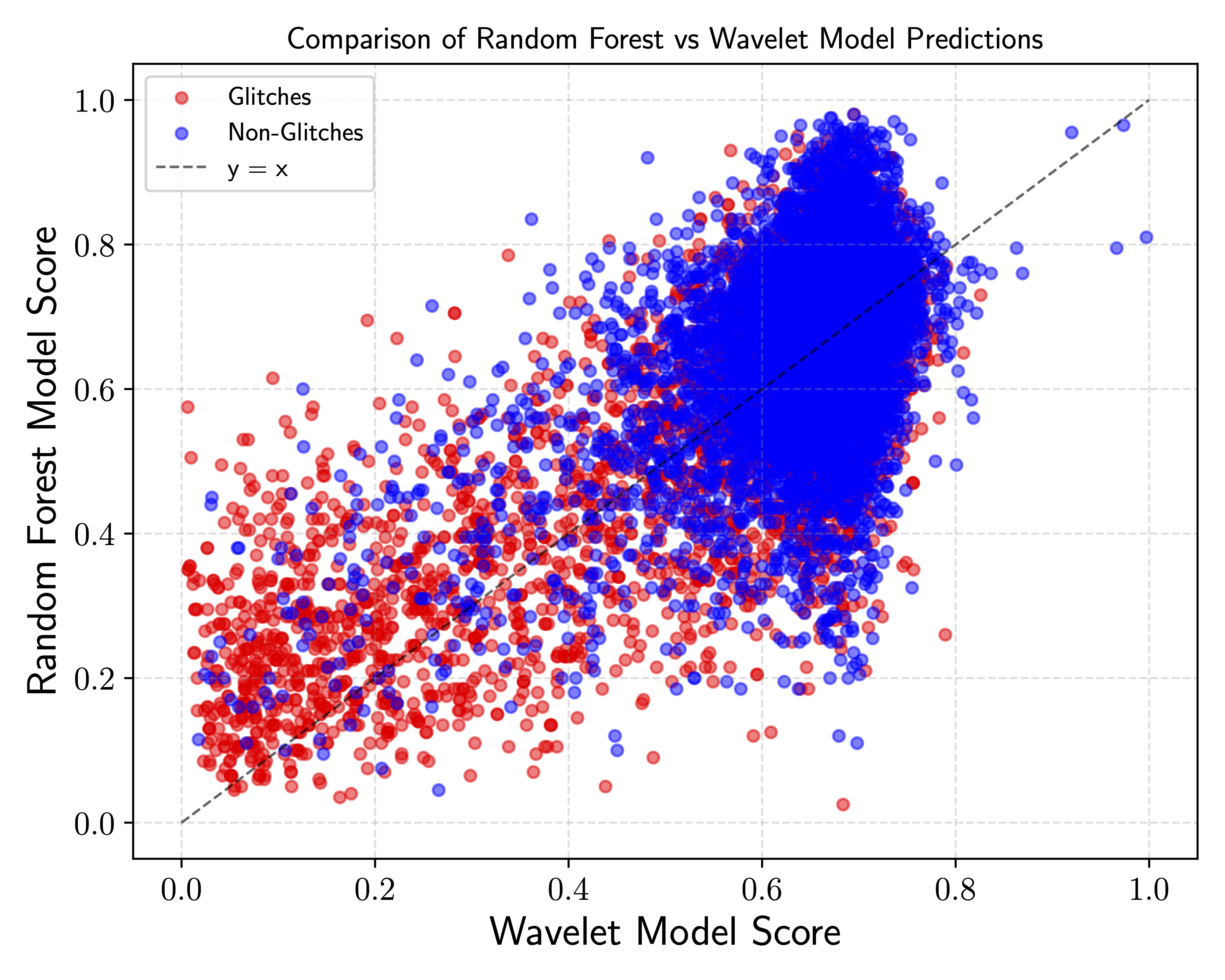}
    \caption{This scatter plot compares the  output scores for the RF model trained in~\cite{Wadekar:2025lhk} and for the  \texttt{WaveletNet} model. The scattering of the points away from the \texttt{y = x} line shows that \texttt{WaveletNet} learns features different from the RF model. }
    \label{fig:scatterplot}
\end{figure}

\begin{figure}[t]
    \centering
    \includegraphics[width=0.9\linewidth]{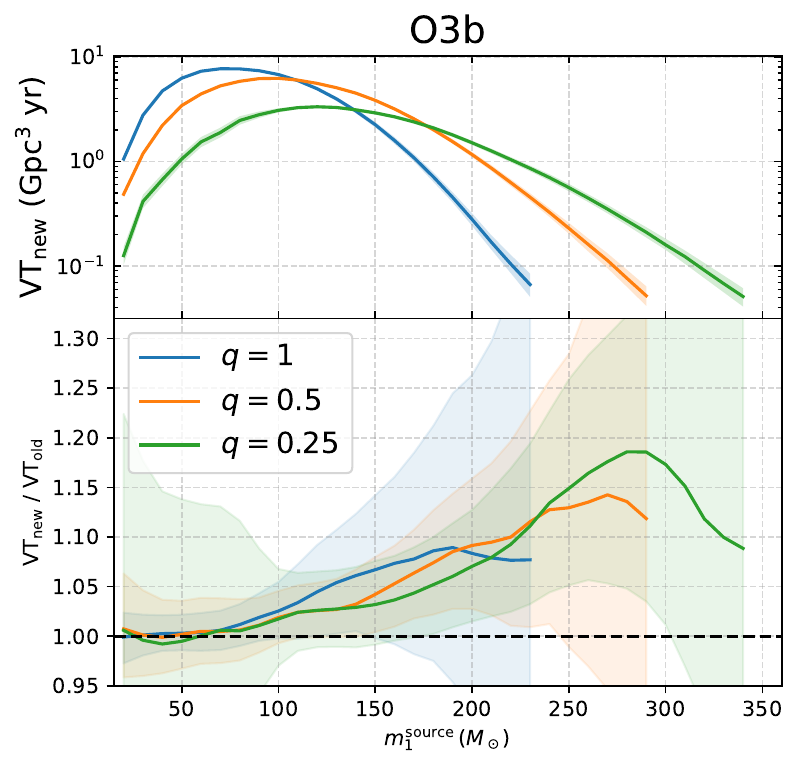}
    \caption{Similar to Fig.~\ref{fig:VT_increase}, but using a RF regression model instead of the multi-layer perceptron as the final classification module (see Fig.~\ref{fig:WaveletNet_workflow}). This comparison tests the choice of final classifier. We find that the resulting increase in $VT$ is comparable to that obtained with the MLP-based \texttt{WaveletNet}.}
    \label{fig:wavelet_RF}
\end{figure}

\begin{figure}[t]
    \centering
    \includegraphics[width=0.9\linewidth]{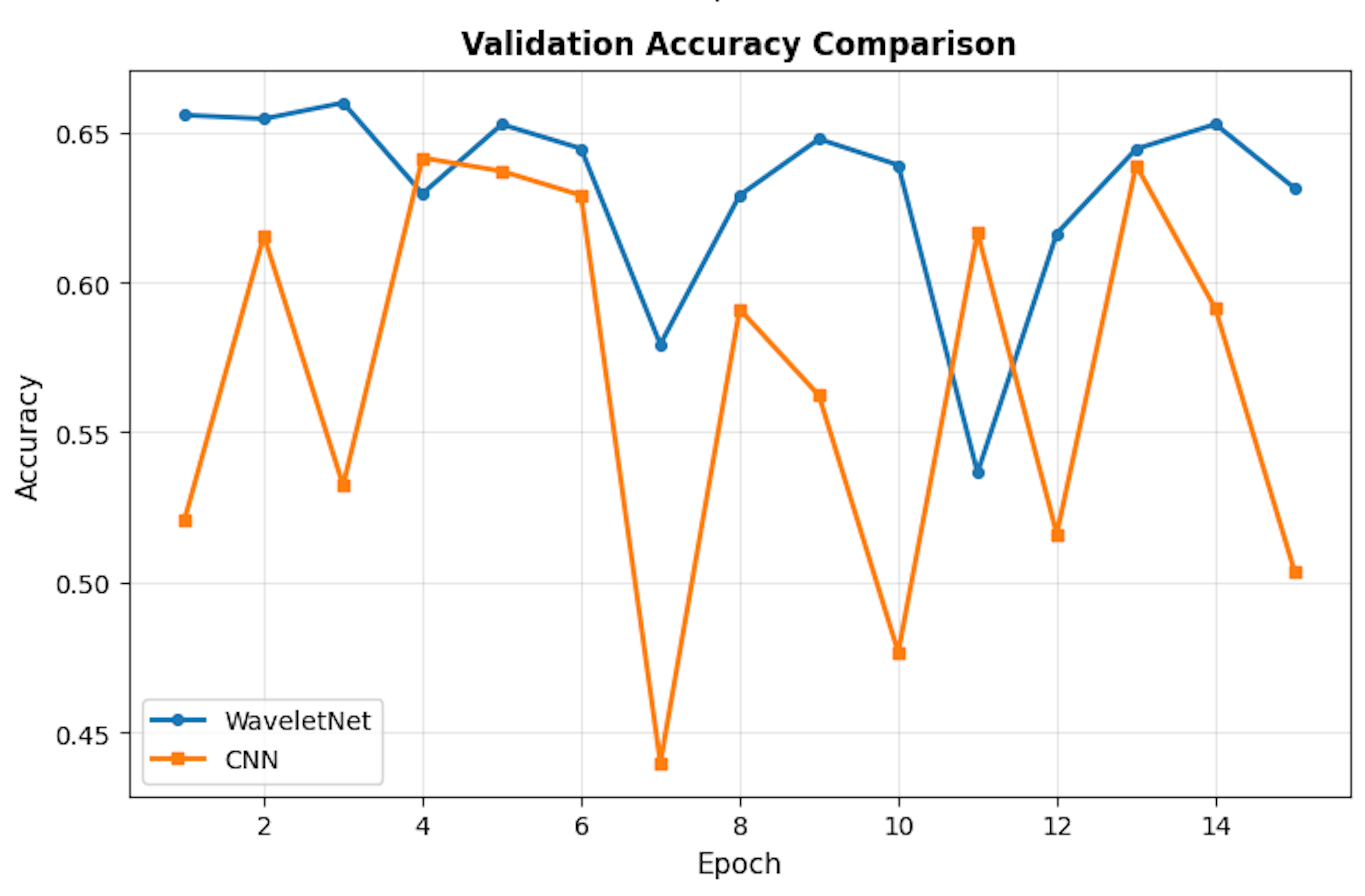}
    \caption{Validation accuracy over training epochs for \texttt{WaveletNet} and for the CNN. \texttt{WaveletNet} consistently achieves higher and more stable performance than the CNN. Training CNNs requires large training datasets, especially when the input data segments are long.}
    \label{fig:cnn_vs_waveletnet_val_loss}
\end{figure}

\begin{figure}[!htbp]
    \centering
        \includegraphics[width=0.9\linewidth]{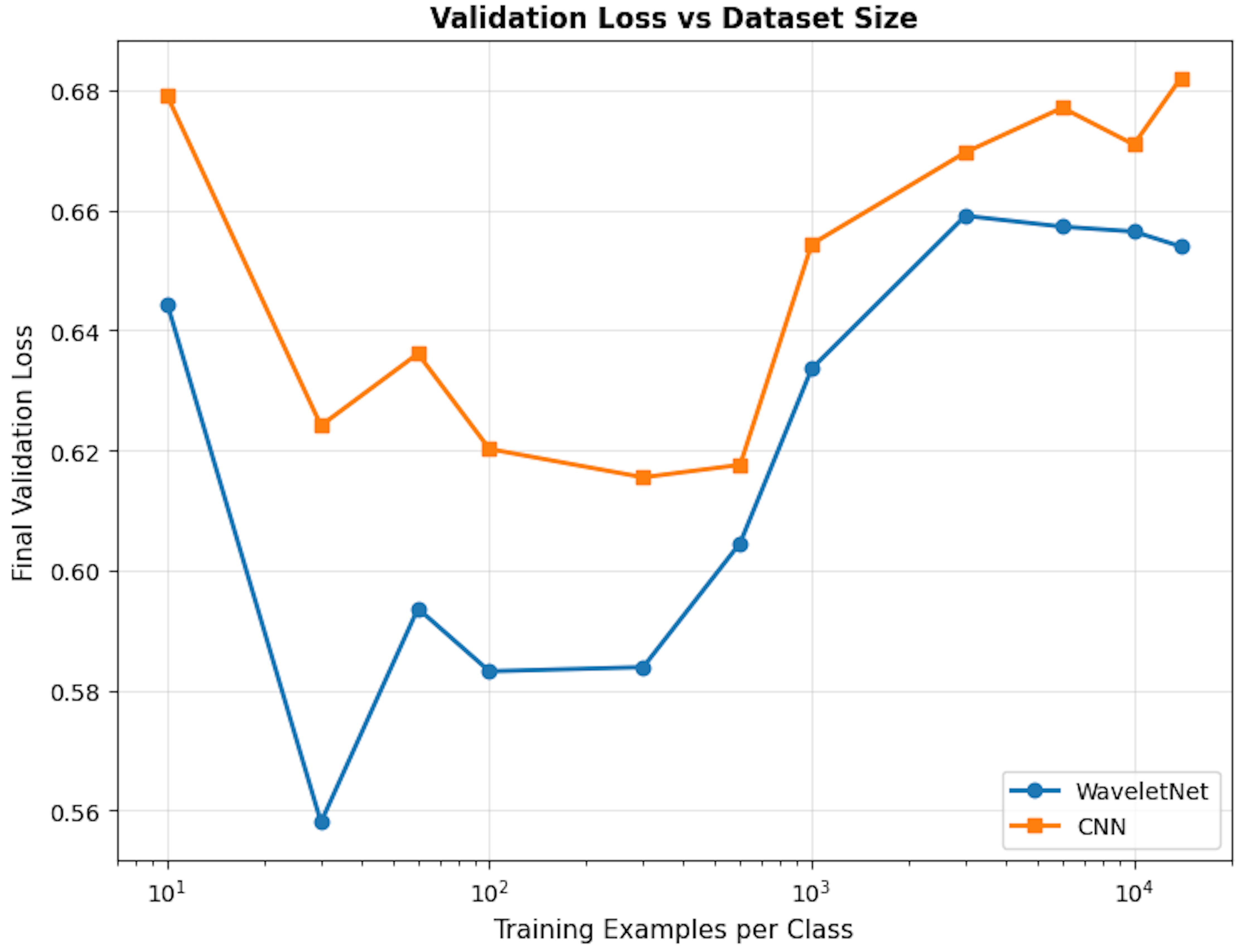}
    \caption{Final validation loss as a function of training set size for \texttt{WaveletNet} and CNN. \texttt{WaveletNet} maintains lower and more stable loss across dataset sizes, while CNN performance degrades at larger scales, reflecting its higher data requirements.}
    \label{fig:val_vs_dataset_size}
\end{figure}

In our previous work~\cite{Wadekar:2025lhk}, we trained a Random Forest (RF) model only on the summary statistics that achieved almost 20\% improvement in VT. The \texttt{WaveletNet} model is a different approach to evaluate the nonlocal environment of candidates. In Fig.~\ref{fig:scatterplot}, we compare the outputs of RF and \texttt{WaveletNet} for glitchy and non-glitchy data strains. We also attempt a variation of \texttt{WaveletNet} by replacing the classification-head MLP with an RF model and plot its VT increase in Fig. \ref{fig:wavelet_RF}. In Figs.  \ref{fig:cnn_vs_waveletnet_val_loss} and \ref{fig:val_vs_dataset_size}, we show that \texttt{WaveletNet} achieves more stable validation performance than CNNs, particularly in the limited-data regime.

\FloatBarrier
\bibliography{wavelet_glitch}   

\end{document}